\documentclass[doublecol]{epl2} 

\usepackage{graphicx}
\usepackage{color}
\usepackage{epsfig}
\usepackage{amssymb}
\usepackage{amsfonts}
\usepackage{amsmath}

\usepackage{multicol}

\twocolumn

\title{Photon emission and atomic collision processes in two-phase \\ argon doped with xenon and nitrogen}

\author{A. Buzulutskov\inst{1,2}}
\shortauthor{A. Buzulutskov}

\institute{
  \inst{1} Budker Institute of Nuclear Physics SB RAS, Novosibirsk, 630090, Russia\\
  \inst{2} Novosibirsk State University, Novosibirsk, 630090, Russia
}

\pacs{95.55.Vj}{Neutrino, muon,pion, and other elementary particle detectors; cosmic ray detectors} \pacs{61.25.Bi}{Liquid noble gases} \pacs{95.35.+d}{Dark matter}

\abstract{We present a comprehensive analysis of photon emission and atomic collision processes in two-phase argon doped with xenon and nitrogen. The dopants are aimed to convert the VUV emission of pure Ar to the UV emission of the Xe dopant in the liquid phase and to the near UV emission of the N$_2$ dopant in the gas phase. Such a mixture is relevant to two-phase dark matter and low energy neutrino detectors, with enhanced photon collection efficiency for primary and secondary scintillation signals. Based on this analysis, we show that the recently proposed hypothesis of the enhancement of the excitation transfer from Ar to N$_2$ species in the two-phase mode is either incorrect or needs assumption about a new extreme mechanism of the excitation transfer coming into force at lower temperatures, in particular that of the resonant excitation transfer via ArN$_2$ compound (van der Waals molecule).}

\begin{document}

\maketitle

\section{Introduction}

Currently two types of two-phase detectors are being used for dark matter search by a number of groups worldwide: those based on liquid xenon \cite{Xenon100,Lux,Panda} and liquid argon \cite{Darkside,Marchionni11}. In addition, the use of argon- and xenon-based two-phase detectors was proposed for coherent neutrino-nucleus scattering experiments \cite{Hagmann04,Akimov10}. A key aspect of two-phase detectors is the simultaneous recording of primary scintillation (S1) and primary ionization (S2) signals \cite{Aprile06,Chepel13,Bernabei15}, providing selection of nuclear recoil events induced by elastic collisions with dark matter particles or low energy neutrinos. Both S1 and S2 signals are optically read out in the liquid and gas phase respectively, the S2 signal being recorded via secondary scintillation (proportional electroluminescence).

In Xe, having a photon emission band in the Ultraviolet (UV), around 175 nm, the S1 and S2 signals are recorded directly: using cryogenic PMTs with quartz windows. In contrast in Ar, having a photon emission band in the Vacuum Ultraviolet (VUV), around 128 nm, the signals are recorded indirectly: using cryogenic PMTs combined with a Wavelength Shifter (WLS), typically TPB \cite{Darkside}, to convert the VUV into the visible range. In this case the photon collection efficiency might be considerably reduced, by a factor reaching 20 \cite{Bondar15},  due internal reflection and conversion efficiency losses in the WLS and in the absence of optical contact between the WLS and the PMT window.

Accordingly, it looks attractive to shift the VUV emission of Ar to longer wavelength directly in the detection medium, to do without the WLS. It is known that at room temperature such a VUV-to-UV conversion can be effectively performed in gaseous argon by doping with either xenon \cite{Gleason77,Takahashi83}, at a content of 0.1-10\%, or nitrogen \cite{Policarpo67,Takahashi83,Kazkaz10}, at a content of 0.2-2\%, and in liquid argon by doping with xenon \cite{Cheshnovsky72,Kubota82,Kubota93,Hitachi93,Wahl14,Neumeier15a,Neumeier15}, at a content of 10-1000 ppm.

In two-phase Ar, doping with Xe is effective only in the liquid phase, since in the gas phase the saturated Xe vapor pressure is as low as 0.031 Torr at 87.3 K \cite{Fastovsky72}, which for example at Xe content of 1000 ppm in the liquid results in vanishing Xe content in the gas phase, of 40 ppb (according to Rault's law). In addition, doping liquid Ar with even minor amount of N$_2$ ($\geq$10 ppm) results in quenching the VUV emission \cite{Himi82,Acciarri10}, without any re-emission in the UV. Consequently at first glance, the VUV-to-UV conversion in two-phase Ar would be possible either for the S1 signal in the liquid phase, using Xe dopant, or for the S2 signal in the gas phase, using N$_2$ dopant, but not for both the S1 and S2 signals.

On the other hand, the recent results on electroluminescence yield in two-phase Ar doped with a rather small amount of N$_2$, at a content of 50 ppm, were explained by a hypothesis that 50\% of emitted photons were due to N$_2$ emission \cite{Bondar15,Bondar17}. This hypothesis implies that the excitation transfer from Ar to N$_2$ species is substantially enhanced at 87 K compared to room temperature. Given such a low N$_2$ content, one could be tempted to use a ternary mixture of argon doped with xenon and nitrogen, in which the N$_2$ dopant in the liquid is supposed to do not interfere with that of Xe.

In this paper, we try to resolve these questions through careful analysis of energy levels, photon emission bands and reaction rate constants of Ar, Xe and N$_2$ species relevant to the performance in the two-phase mode.

The present study was performed in the course of the development of two-phase Cryogenic Avalanche Detectors (CRADs) of ultimate sensitivity for rare-event experiments \cite{Buzulutskov12,Bondar14,Bondar16,CryoGAPD}.

\section{Ternary mixture of Ar doped with Xe and N$_2$}

As a specific example, we examine here a hypothetical ternary mixture of Ar doped with Xe and N$_2$, at a content of 1000 ppm and 50 ppm in the liquid and 40 ppb and 135 ppm in the gas phase, respectively. The dopant contents in the gas phase were defined according to Rault's law, from the saturated vapor pressure data at 87 K \cite{Fastovsky72}. The data on energy levels, photon emission bands and reaction rate constants of Ar, Xe and N$_2$ species are compiled here over the past 50 years.

Fig. 1 shows the data relevant to this mixture and to the two-phase detector performance, i.e. when the excitation is due to ionization or electroluminescence: the energy levels of the lower excited and ionized states, the radiative transitions observed in experiments and the most probable non-radiative transitions induced by atomic collisions for Ar, Xe and N$_2$ species and their pair combinations (Ar+Xe and Ar+N$_2$). Table 1 explains and elaborates Fig.1, presenting basic reactions of excited species for a given mixture, relevant to the performance in the two-phase mode, their rate ($k$) or time ($\tau$) constants reported in the literature and their time constants reduced to given atomic densities at 87 K ($\tau_{TP}$).

The latter constants allow to rank the reactions on importance in case of parallel competing reactions: the smaller the time constant of the reaction, the larger its contribution. When calculating these constants, the second and third order reactions were reduced to pseudo-first order reactions, because the second and third species are present here in large excess with respect to the excited ones \cite{Atkins78}. For example, the reduced time constant of reaction (1) is $\tau_{TP}=1/(k_1 [Ar]^2)$, while that of reaction (6)  is $\tau_{TP}=1/(k_6 [N_2])$.

Below an overview of the reactions of Table 1 is given. In gaseous Ar, there are four lowest excited atomic states Ar$^{\ast}(3p^54s^1)$: those of two resonance ($^3P_1$ and $^1P_1$) and two  meta-stable ($^3P_2$ and $^3P_0$). These excited states and three-body collision reaction (1) are responsible for the Ar$_2^{\ast}$ excimer production in a singlet ($^{1}\Sigma^+_u$) or triplet ($^{3}\Sigma^+_u$) state \cite{Sauerbrey81,Smirnov83,Krylov02,Lorents76}, followed by their radiative decay in the VUV, at 128$\pm$12 nm (reaction (2)) \cite{Cheshnovsky72,Lorents76,Keto74,Policarpo81,Gleason77,Suzuki82,Monteiro08}. The singlet and triplet states provide the fast and slow emission component respectively. Besides the VUV emission, the excited Ar atoms in the gas phase emit in the Near Infrared (NIR), at 690-850 nm, due to the higher excited atomic states Ar$^{\ast}(3p^54p^1)$ (reaction (3)) \cite{Lindblom88,Hofmann13,Wiese89,Oliveira13,Buzulutskov11,Bondar12,Bondar12a}. Due to the large uncertainty in the reaction (1) rate constant measured at room temperature, its reduced time constant at 87 K is defined only within an order of magnitude: $\tau_{TP}$$\sim$13 ns. Here and below the weak temperature dependencies of the rate constants \cite{Smirnov83} were neglected.

Despite the relatively large rate constants of reactions (4) and (5), describing the excitation transfer from Ar$^{\ast}$ and Ar$_2^{\ast}$ states to Xe \cite{Velazco78,Takahashi83,Oka79,Gleason77}, their contributions are negligible due to the extremely low Xe content in the gas phase: compare their $\tau_{TP}$ to those of reactions (1) and (2).

\section{N$_2^{\ast}(C)$ versus Ar$_2^{\ast}$ emission in the gas phase}

Reaction (6) is responsible for the VUV-to-UV conversion in the gas phase: it describes two-body collisions of Ar$^{\ast}$ states with N$_2$ molecules, producing N$_2^{\ast}(C^3\Pi_u)$ excited states (N$_2^{\ast}(C)$ for short) \cite{Takahashi83,Sauerbrey81,Krylov02,Sadeghi81}; it competes with reaction (1). Here the rate constant at room temperature is known within a factor of 2 \cite{Sauerbrey81,Krylov02,Sadeghi81,Velazco78}, resulting in that the N$_2^{\ast}(C)$ fraction of all the Ar+N$_2$ reaction products may range from 50 to 100\% (compare to reaction (7)). Reaction (6) is followed by reaction (8): by N$_2^{\ast}(C)$$\rightarrow$N$_2^{\ast}(B)$ radiative transition in the near UV, emitting the so-called 2nd Positive System (2PS) at 310-440 nm \cite{Takahashi83,Sauerbrey81,Millet73,Krylov02}.

According to its rate constant measured at room temperature, reaction (6) would not be able to compete with reaction (1) at such a small (135 ppm) N$_2$ content: its $\tau_{TP}$=2.4 $\mu$s would be at least a factor of 180 larger than that of reaction (1). What confuses is that recently the opposite thing has been reported \cite{Bondar15,Bondar17}, namely that 50\% of photons produced due to Ar$^{\ast}$ species in the gas phase at 87 K might be emitted by those of N$_2^{\ast}(C)$, at N$_2$ content in the liquid of 50 ppm. This was explained by the hypothesis that the excitation transfer from Ar to N$_2$ is enhanced at 87 K compared to room temperature.

Given the analysis performed, we tend to believe that such hypothesis can hardly be true, leading us to the conclusion that the experimental data of \cite{Bondar15} are incorrect. Otherwise one should assume an extreme mechanism of atomic collisions at which the rate constant of reaction (6) at 87 K is increased by a huge factor ($\geq$180), compared to that of room temperature; below, in the last-but-one section, we will examine this possibility.

The excited N$_2^{\ast}(C)$ states can be quenched in collisions with Ar and N$_2$ species in reaction (10) and (12); in particular reaction (10) competes with reaction (8), somewhat limiting the photon yield of the 2PS emission.

\onecolumn

\begin{multicols}{2}

Reaction (9) can follow reaction (8), (10) and (12): it describes N$_2^{\ast}(B)$$\rightarrow$N$_2^{\ast}(A)$ radiative transition in the NIR, emitting the 1st Positive System (1PS) at 500-2500 nm \cite{Friedl12,Boesch15,Sauerbrey81,Berg94}. At normal and higher pressures  the 1PS emission is suppressed due to its large time constant (9 $\mu$s) and quenching reactions (11) and (13). For the same reason, we do not consider here the N$_2^{\ast}(A)$$\rightarrow$N$_2(X)$ radiative transition, since its time constant is as large as several ms \cite{Berg94}.

Reaction (14) is similar to reaction (5); it describes quenching of Ar$_2^{\ast}$ species in collisions with those of N$_2$. In the gas phase its contribution is inessential due to the large time constant. In contrast in the liquid phase, it may play an important role in quenching the VUV emission (see below reaction (23)).

\section{Xe$_2^{\ast}$ emission versus N$_2$-induced quenching in the liquid phase}

In the liquid, the excited, ground and ionized atomic states transform to the exciton, valence and conduction bands, respectively (see Fig. 1): in particular in Ar, to the Ar$^{\ast}(n=1,{}^2P_{3/2})$ and Ar$^{\ast}(n=1,{}^2P_{1/2})$ excitons \cite{Schwentner85}, which reflect the ${}^3P_1$ and ${}^1P_1$ levels of the Ar$^{\ast}(3p^54s^1)$ atomic states. According to reaction (15), the excitons are immediately trapped in singlet or triplet excimer states Ar$_2^{\ast}(^{1,3}\Sigma^+_u)$ \cite{Martin71,Acciarri10}. These states have lower energy levels with respect to the  N$_2^{\ast}(C)$ states (see Fig. 1); that is why the 2PS emission in the UV had never been observed in liquid Ar doped with N$_2$. The singlet ($^{1}\Sigma^+_u$) and triplet ($^{3}\Sigma^+_u$) excimers provide the fast and slow emission components in the VUV (reaction (16)).
\end{multicols}

\begin{figure}
\includegraphics[width=0.99\columnwidth]{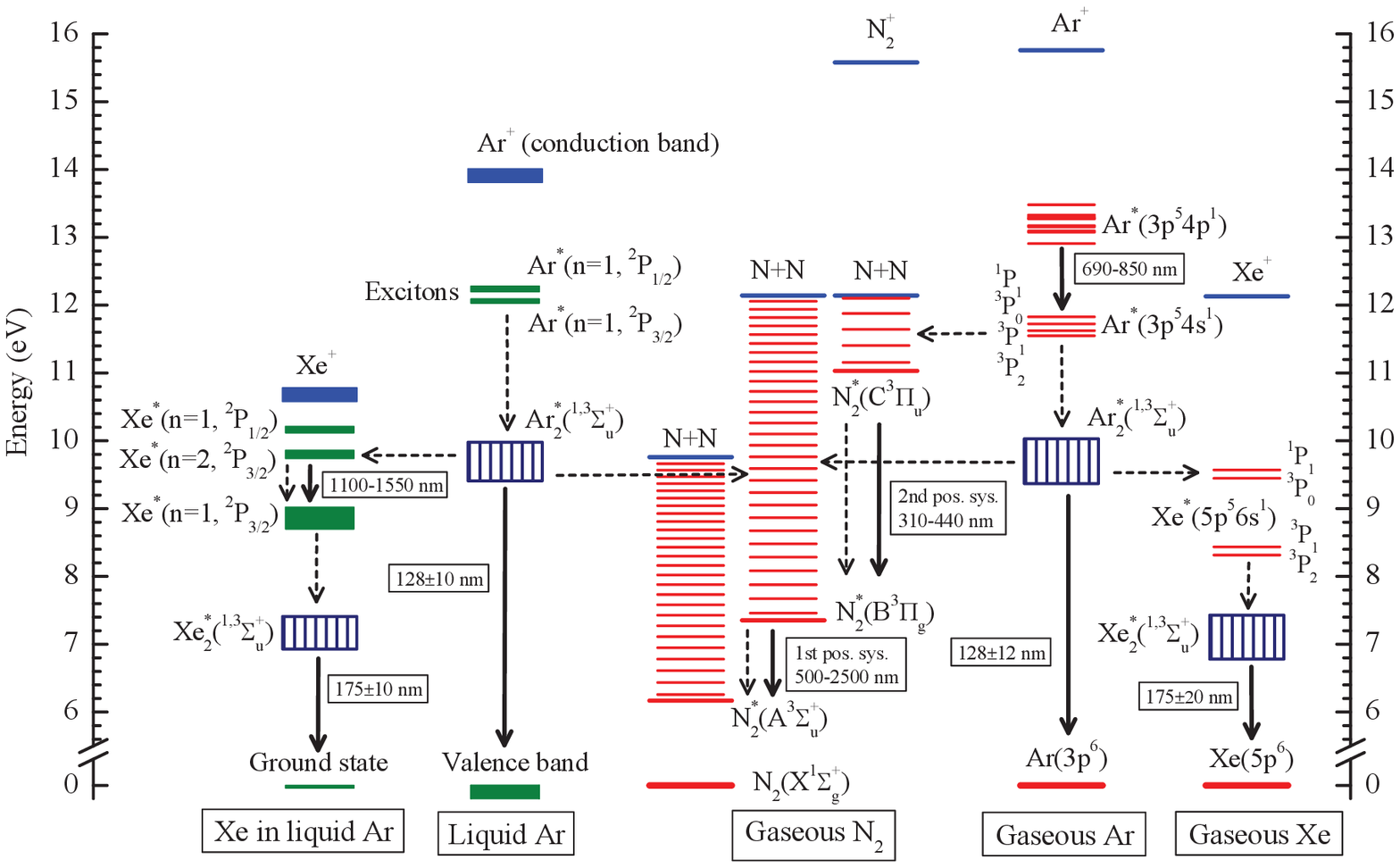}
\caption{Energy levels of the lower excited and ionized states relevant to the ternary mixture of Ar doped with Xe and N$_2$ in the two-phase mode; these are shown for gaseous Ar \cite{NIST}, gaseous N$_2$ \cite{Lofthus77,Radtsig80}, gaseous Xe \cite{NIST}, liquid Ar \cite{Schwentner85,Jortner77,Raz70,Hitachi93,Badhrees10} and Xe in liquid Ar \cite{Neumeier15,Raz70}. For N$_2$, the vibrational levels of the $C^3\Pi_u$, $B^3\Pi_g$ and $A^3\Sigma^+_u$ molecular states, as well as the dissociation levels (N+N), are also shown (except those of the ground state $X^1\Sigma^+_g$). Due to the lack of data for liquid Ar, its exciton and valence bands are taken the same as for solid Ar \cite{Schwentner85}. The solid arrows indicate the radiative transitions observed in experiments and relevant to the present study (i.e. when the excitation is induced by ionization or electroluminescence): Ar$_2^{\ast}$ in gaseous and liquid Ar \cite{Cheshnovsky72}, Xe$_2^{\ast}$ in gaseous Xe \cite{Jortner65}, Xe$_2^{\ast}$ in liquid Ar \cite{Cheshnovsky72}, Ar$^{\ast}$ in gaseous Ar in the NIR \cite{Lindblom88,Hofmann13}, N$_2^{\ast}$ in gaseous Ar+N$_2$ in the UV (2nd positive system) \cite{Takahashi83}, N$_2^{\ast}$ in gaseous Ar+N$_2$ in the NIR (1st positive system) \cite{Friedl12,Boesch15,Berg94} and Xe$^{\ast}$  in liquid Ar in the NIR \cite{Neumeier15,Neumeier15a}.  The numbers next to each arrow show the photon emission band of the transition, defined by major emission lines or by full width of the emission continuum. In addition, for transitions between the Ar$_2^{\ast}$ and Xe$_2^{\ast}$ excimers and the ground states, the full width at half-maximum (FWHM) of the emission continuum is denoted by the vertically shaded area. The dashed arrows indicate the most probable non-radiative transitions induced by atomic collisions for Ar, Xe and N$_2$ species and their pair combinations in the gas and liquid phase.}
\label{Levels}
\end{figure}

\begin{table}
\caption{Basic reactions of excited species relevant to the performance in the two-phase mode, namely in Ar in the gas and liquid phase, doped with Xe (1000 ppm in the liquid and 40 ppb in the gas phase) and N$_2$ (50 ppm in the liquid and 135 ppm in the gas phase),  their rate ($k$) or time ($\tau$) constants reported in the literature and their time constants reduced to given atomic densities at 87 K ($\tau_{TP}$), in particular for Ar to that of 8.63$\times$10$^{19}$ cm$^{-3}$ and 2.11$\times$10$^{22}$ cm$^{-3}$ in the gas and liquid phase respectively.}
\label{tab.1}
\begin{center}
\begin{tabular}{llllll}
No. & Reaction                                                                      & $k$ or $\tau$ & T & Reference & $\tau_{TP}$ \\
[2mm]\hline
\\   & \multicolumn{2}{l}{\underline{Gaseous Ar + Xe (40 ppb) + N$_2$ (135 ppm)}} \\
(1)  & \( Ar^{\ast}(3p^54s^1) + 2Ar \rightarrow \)                                  & $k_1$$\sim$1$\times$10$^{-32}$ cm$^6$ s$^{-1}$  & 300 K & \cite{Sauerbrey81,Smirnov83,Krylov02,Lorents76} & $\sim$13 ns \\
     & \( \rightarrow Ar_2^{\ast}(^{1,3}\Sigma^+_u) + Ar \)                         & \\
(2)  & \( Ar_2^{\ast}(^{1,3}\Sigma^+_u) \rightarrow 2Ar + h\nu \, (\mathrm{VUV}) \) & $\tau_2(^{1}\Sigma^+_u)$=4.2 ns  & 300 K & \cite{Keto74,Policarpo81} & 4.2 ns \\
     &                                                                              & $\tau_2(^{3}\Sigma^+_u)$=3.0-3.2 $\mu$s & 300 K & \cite{Lorents76,Keto74,Policarpo81,Gleason77,Suzuki82,Monteiro08} & 3.1 $\mu$s \\
(3)  & \( Ar^{\ast}(3p^54p^1) \rightarrow \)                                        & $\tau_3$=20-40 ns & 300 K & \cite{Lindblom88,Wiese89,Oliveira13}  \\
     & \( \rightarrow Ar^{\ast}(3p^54s^1) + h\nu \, (\mathrm{NIR}) \)               & $\tau_3$$<$100 ns & 163 K & \cite{Buzulutskov11,Bondar12,Bondar12a}  & $<$100 ns \\
(4) & \( Ar^{\ast}(3p^54s^1) + Xe \rightarrow Ar + Xe^{\ast} \)                     & $k_4$=(2-3)$\times$10$^{-10}$ cm$^3$ s$^{-1}$  & 300 K & \cite{Velazco78,Takahashi83} & $\sim$1 ms \\
(5) & \( Ar_2^{\ast}(^{3}\Sigma^+_u) + Xe \rightarrow \)                            & $k_5$$\sim$5$\times$10$^{-10}$ cm$^3$ s$^{-1}$  & 300 K & \cite{Oka79,Gleason77,Takahashi83} & $\sim$0.6 ms \\
     & \( \rightarrow 2Ar + Xe^{\ast}({}^1P_1,{}^3P_0) \) \\
(6)  & \( Ar^{\ast}(3p^54s^1) + N_2 \rightarrow Ar + N_2^{\ast}(C) \)               & $k_6$$\sim$1.5$\times$10$^{-11}$ cm$^3$ s$^{-1}$  & 300 K & \cite{Sauerbrey81,Krylov02} & \\
     &                                                                              & $k_6$=3.6$\times$10$^{-11}$ cm$^3$ s$^{-1}$  & 300 K & \cite{Sadeghi81} & 2.4 $\mu$s \\
     &                                                                              & $k_6$$\geq$6.5$\times$10$^{-9}$ cm$^3$ s$^{-1}$ (?) & 87 K & \cite{Bondar15} & $\leq$13 ns (?) \\
(7)  & \( Ar^{\ast}(3p^54s^1) + N_2 \rightarrow \)                                  & $k_7$$\sim$3$\times$10$^{-11}$ cm$^3$ s$^{-1}$  & 300 K & \cite{Sauerbrey81,Krylov02} \\
     & \( \rightarrow Ar + N_2^{\ast}(C,B,A) \)                                     & $k_7$=3.6$\times$10$^{-11}$ cm$^3$ s$^{-1}$  & 300 K & \cite{Sadeghi81,Velazco78} \\
(8)  & \( N_2^{\ast}(C) \rightarrow \)                                              & $\tau_8$=30-40 ns& 300 K & \cite{Sauerbrey81,Millet73,Krylov02} & 35 ns \\
     & \( \rightarrow N_2^{\ast}(B) + h\nu \, (\mathrm{UV, 2nd \, pos. \, sys.}) \) \\
(9) & \( N_2^{\ast}(B) \rightarrow \)                                               & $\tau_{9}$$\sim$9 $\mu$s & 300 K & \cite{Sauerbrey81} & $\sim$9 $\mu$s \\
     & \( \rightarrow N_2^{\ast}(A) + h\nu \, (\mathrm{NIR, 1st \, pos. \, sys.}) \)& & 119 K & \cite{Berg94} \\
(10) & \( N_2^{\ast}(C) + Ar \rightarrow N_2^{\ast}(B) + Ar \)                      & k$_{10}$=5.6$\times$10$^{-13}$ cm$^3$ s$^{-1}$ & 300 K & \cite{Sauerbrey81} & 21 ns \\
(11) & \( N_2^{\ast}(B) + Ar \rightarrow N_2^{\ast}(A) + Ar \)                      & k$_{11}$=1.4$\times$10$^{-14}$ cm$^3$ s$^{-1}$ & 300 K & \cite{Sauerbrey81} & 0.8 $\mu$s \\
(12)  & \( N_2^{\ast}(C) + N_2 \rightarrow N_2+ N_2^{\ast}(B) \)                    & $k_{12}$$\sim$1$\times$10$^{-11}$ cm$^3$ s$^{-1}$ & 300 K & \cite{Sauerbrey81,Millet73} & $\sim$8.6 $\mu$s \\
(13) & \( N_2^{\ast}(B) + N_2 \rightarrow N_2 + N_2^{\ast}(A) \)                    & $k_{13}$$\sim$1$\times$10$^{-11}$ cm$^3$ s$^{-1}$ & 300 K & \cite{Sauerbrey81} & $\sim$8.6 $\mu$s \\
(14) & \( Ar_2^{\ast}(^{3}\Sigma^+_u) + N_2 \rightarrow 2Ar + N_2^{\ast}(B) \)      & $k_{14}$$\sim$3.3$\times$10$^{-12}$ cm$^3$ s$^{-1}$  & 300 K & \cite{Sauerbrey81,Oka79} & $\sim$26 $\mu$s \\

\\ & \multicolumn{2}{l}{\underline{Liquid Ar + Xe (1000 ppm) + N$_2$ (50 ppm)}} \\
(15) & \( Ar^{\ast}(n=1,{}^2P_{1/2,3/2}) + Ar \rightarrow \)                        & $\tau_{15}$=6 ps & 87 K & \cite{Martin71,Acciarri10} & 6 ps \\
     & \( \rightarrow Ar_2^{\ast}(^{1,3}\Sigma^+_u) \) \\
(16) & \( Ar_2^{\ast}(^{1,3}\Sigma^+_u) \rightarrow 2Ar + h\nu \, (\mathrm{VUV}) \) & $\tau_{16}(^{1}\Sigma^+_u)$=7 ns & 87 K & \cite{Hitachi83,Aprile06,Chepel13,Bernabei15}  & 7 ns \\
     &                                                                              & $\tau_{16}(^{3}\Sigma^+_u)$=1.6 $\mu$s & & & 1.6 $\mu$s \\
(17) & \( Ar_2^{\ast}(^{1,3}\Sigma^+_u) + Xe \rightarrow \)                         & $k_{17}(^{3}\Sigma^+_u)$$\sim$ & 87 K & \cite{Kubota82,Kubota93,Hitachi93} & $\sim$5.3 ns \\
     & \( \rightarrow 2Ar \ + Xe^{\ast}(n=1,2,{}^2P_{3/2}) \)                       & $\sim$(0.8-1)$\times$10$^{-11}$ cm$^3$ s$^{-1}$ \\
     &                                                                              & $\tau_{17}(^{3}\Sigma^+_u)$$<$90 ns & 87 K & \cite{Kubota93,Wahl14} & $<$90 ns \\
     &                                                                              & $k_{17}(^{1}\Sigma^+_u)$$\sim$3.3$\times$10$^{-11}$ cm$^3$ s$^{-1}$ & 87 K & \cite{Hitachi93} & $\sim$1.4 ns \\
(18) & \( Xe^{\ast}(n=1,2,{}^2P_{3/2}) + Ar \rightarrow ArXe^{\ast} \)              & Immediate trapping & 87 K & \cite{Hitachi93}\\
(19) & \( ArXe^{\ast} + Xe \rightarrow Ar + Xe_2^{\ast}(^{1,3}\Sigma^+_u) \)        & $\tau_{19}$$\le$20 ns & 87 K & \cite{Wahl14} & $\le$20 ns \\
(20) & \( Xe^{\ast}(n=1,2,{}^2P_{3/2}) + Xe \rightarrow \)                          & - & 87 K & \cite{Cheshnovsky72} \\
     & \( \rightarrow Xe_2^{\ast}(^{1,3}\Sigma^+_u) \)  \\
(21) & \( Xe_2^{\ast}(^{1,3}\Sigma^+_u) \rightarrow 2Xe + h\nu \, (\mathrm{UV}) \)  & $\tau_{21}(^{1}\Sigma^+_u)$=4.3 ns & 165 K & \cite{Hitachi83,Chepel13} & 4.3 ns \\
     &                                                                              & $\tau_{21}(^{3}\Sigma^+_u)$=22 ns & 165 K & & 22 ns \\
(22) & \( Xe^{\ast}(n=2,{}^2P_{3/2}) \rightarrow \)                                 & $\tau_{22}<$170 ns & 87 K & \cite{Neumeier15,Neumeier15a} & $<$170 ns \\
     & \( \rightarrow Xe^{\ast}(n=1,{}^2P_{3/2}) + \ h\nu \, (\mathrm{NIR}) \) \\
\\   & \multicolumn{2}{l}{Reactions (17)-(21) in total ($\tau_{17}$+$\tau_{19}$):}  & & & $\leq$110 ns \\ \\
(23) & \( Ar_2^{\ast}(^{3}\Sigma^+_u) + N_2 \rightarrow 2Ar + N_2^{\ast}(B) \)      & $k_{23}$=3.8$\times$10$^{-12}$ cm$^3$ s$^{-1}$ & 87 K & \cite{Himi82,Acciarri10}  & 250 ns \\
(24) & \( ArXe^{\ast} + N_2 \rightarrow Ar + Xe + N_2^{\ast}(B,A) \)                  & - & 87 K & \ &  \\
(25) & \( Xe_2^{\ast}(^{3}\Sigma^+_u) + N_2 \rightarrow 2Xe + N_2^{\ast}(B,A) \)       & - & 87 K & \ &  \\

\end{tabular}
\end{center}
\end{table}

\twocolumn

Reactions (17)-(21), describing the excitation transfer from Ar$_2^{\ast}(^{1,3}\Sigma^+_u)$ excimers to those of Xe$_2^{\ast}(^{1,3}\Sigma^+_u)$, are the major reactions responsible for the VUV-to-UV conversion in the liquid phase due to Xe dopant.  At the first stage, Xe excitons are produced in excitation transfer from Ar$_2^{\ast}$ excimers in reaction (17) \cite{Kubota82,Kubota93,Hitachi93,Wahl14}; it competes with reaction (16). Similarly to Ar, the Xe$^{\ast}(n=1,{}^2P_{3/2})$ and Xe$^{\ast}(n=1,{}^2P_{1/2})$ excitons in liquid Ar reflect the ${}^3P_1$ and ${}^1P_1$ levels of the Xe$^{\ast}(5p^56s^1)$ atomic states; in addition, a Wannier-Mott exciton Xe$^{\ast}(n=2,{}^2P_{3/2})$ locates in-between \cite{Schwentner85,Neumeier15,Raz70}. The radiative transition between the latter and the lower exciton band (reaction (22)) is responsible for the NIR emission observed at a smaller Xe content ($<$100 ppm) \cite{Neumeier15,Neumeier15a}.

The time constant of reaction (17), 1.4 ns \cite{Kubota82,Kubota93,Hitachi93} and 5-90 ns \cite{Kubota93,Wahl14} for collisions with the singlet and triplet Ar$_2^{\ast}$ excimers respectively, make this reaction 5-20 times faster than reaction (16) (compare to $\tau_{16}$ in Table 1), giving the latter no chance to happen.

At the second stage, the Xe$^{\ast}$ excitons transform to Xe$_2^{\ast}(^{1,3}\Sigma^+_u)$ excimers in two ways: either via intermediate heteronuclear excimers ArXe$^{\ast}$, in successive collisions with Ar and Xe atoms (reactions (18) and (19)) \cite{Kubota93,Wahl14,Hitachi93}, or directly in collisions with Xe atoms (reaction (20)) \cite{Cheshnovsky72}. The first way is preferred since the  Xe$^{\ast}$ exciton trapping in collisions with Ar atoms (reaction (18)) is believed to occur as fast as reaction (15), i.e. practically immediately \cite{Hitachi93}. On the other hand, the second way cannot be fully excluded for a smaller Xe content ($<$100 ppm), since reaction (22) may indicate on the presence of free-migrating Xe$^{\ast}$ species. Anyway, the rate and time constants measured for reactions (17)-(19) \cite{Kubota93,Wahl14,Hitachi93} can be used here to estimate the overall time for reactions (17)-(21): it amounts to $\leq$110 ns.

This should be compared to the time constants of competitive reactions (23)-(25), describing the non-radiative quenching of Ar$_2^{\ast}$, ArXe$^{\ast}$ and  Xe$_2^{\ast}$ excimers in collisions with N$_2$ molecules. The rate constants of the two latter reactions are assumed to be close to that of (23), which was measured with rather good accuracy \cite{Himi82,Acciarri10}, resulting in the reaction time constant of 250 ns. According to kinetics equation for two parallel reactions \cite{Atkins78}, the quenching reactions contributions would be thus less than (1/$\tau_{23}$)/(1/$\tau_{23}$+1/$\tau_{17-21}$)=30\%.

Accordingly, one may conclude that the Xe dopant to liquid argon at a content of 1000 ppm may successfully perform its job on VUV-to-UV conversion with an efficiency of $>$70\%, even in presence of N$_2$ impurity at a content of $\leq$50 ppm.

\section{On the resonant excitation transfer from Ar to N$_2$}

The hypothesis of the enhancement of the excitation transfer from Ar to N$_2$ species in the gas phase at 87 K \cite{Bondar15} can only be explained if we assume a new mechanism of atomic collisions, namely a resonance behavior of Ar+N$_2$ collision cross-section induced by emergence of a certain bound state around 87 K. It is amazing that such a bound state was actually observed as ArN$_2$ compound (van der Waals molecule) \cite{Henderson74,Smirnov84}, emerging at temperatures below 105 K with a binding energy of 9.6 meV (which is very close to the thermal energy at 87 K). Thus we might suppose that reaction (6) may proceed at 87 K via ArN$_2$ compound in the following resonant reactions:\\
(I) \( \ Ar^{\ast} + ArN_2 \rightarrow Ar + Ar^{\ast}N_2 \ , \)

\( \ \ Ar^{\ast}N_2 \rightarrow ArN_2^{\ast}(C) \rightarrow Ar + N_2^{\ast}(C) \ ; \) \\
(II) \( \ Ar^{\ast} + N_2 \rightarrow Ar^{\ast}N_2 \rightarrow ArN_2^{\ast}(C) \rightarrow Ar + N_2^{\ast}(C) \ .\)

These might considerably enhance the excitation transfer from Ar to N$_2$. In particular, the cross-section of the first reaction might be very large since it includes a resonant excitation transfer reaction $Ar^{\ast} + Ar \rightarrow Ar + Ar^{\ast}$, which is known to have an extremely large cross-section, reaching $\sigma$$\sim$10$^{-12}$ cm$^2$ \cite{Nikitin78}. The rate constant derived hereof is $k$=$\sigma$$\overline{v}$=3$\times$10$^{-8}$ cm$^3$ s$^{-1}$ ($\overline{v}$ is the average relative velocity).

Now the problem is reduced to a question how big the fraction of N$_2$ species bound in ArN$_2$  compound is. It is easy to show that to be consistent with the estimates for reaction (6) deduced from ref. \cite{Bondar15}   and presented in Table 1, $k_6$$\geq$6.5$\times$10$^{-9}$ cm$^3$ s$^{-1}$, this fraction should exceed 20\%. On the other hand, our theoretical estimation showed that this value should be a factor of 20 smaller: $[ArN_2]/[N_2]$$\sim$1\%. Here we used the theoretical calculations of the chemical equilibrium coefficient for Ar$_2$ \cite{Stogryn59} and ArH$_2$ \cite{Hobza88} compounds (defined as $k_{eq}=[ArN_2]/[N_2][Ar]$), given its temperature and binding energy dependence  \cite{Brahms11} to extrapolate to the ArN$_2$ case.

Finally we conclude that the interpretation of the experimental data in \cite{Bondar15} is most probably incorrect and that the effect of the resonant excitation transfer from Ar to N$_2$,  if exists, starts at a much higher N$_2$ content, of about 1000 ppm.

\section{Conclusion}

In this work, a comprehensive analysis of energy levels, photon emission bands and reaction rate constants in two-phase argon doped with xenon and nitrogen, at a content of 1000 ppm and 50 ppm respectively, has been presented. Such a mixture is relevant to two-phase dark matter and low energy neutrino detectors, with enhanced photon collection efficiency for primary and secondary scintillation signals.

Based on this analysis, it is shown that Xe dopant may successfully perform its job on VUV-to-UV conversion in the liquid phase even in presence of N$_2$ impurity, if its content does not exceed 50 ppm.

On the other hand, the recently proposed hypothesis of the enhancement of the excitation transfer from Ar to N$_2$ \cite{Bondar15} which was supposed to provide  the VUV-to-UV conversion in the gas phase, is most probably incorrect at such a small N$_2$ content. We showed that if even to assume a new  extreme mechanism of excitation transfer coming into force at lower temperatures, namely the resonant excitation transfer via ArN$_2$ compound (van der Waals molecule), it presumably starts at a much higher N$_2$ content, of the order of 1000 ppm.

This study was supported by Russian Science Foundation (project N 16-12-10037); it was done within the R\&D program for the DarkSide20k experiment.

\end{document}